\documentclass{CSML}

\pdfoutput=1

\usepackage{lastpage}

\def\dOi{13(4:21)2017}
\lmcsheading%
{\dOi}
{1--\pageref{LastPage}}
{}
{}
{Feb.~24, 2016}
{Nov.~29, 2017}
{}

\usepackage[hidelinks]{hyperref}
\usepackage[utf8]{inputenc}

\theoremstyle{plain}

\begin{document}

\title[From Logic to Biology via Physics: a survey]{From Logic to Biology via Physics: a survey\rsuper*}
 
\author[G.~Longo]{Giuseppe Longo}	
\address{Centre Cavaillès, République des Savoirs,
CNRS, Collège de France et Ecole Normale Supérieure, Paris,
and Department of Integrative Physiology and Pathobiology, 
Tufts University School of Medicine, Boston.}	
\email{giuseppe.longo@ens.fr}  

\author[M.~Montévil]{Maël Montévil}	
\address{IHPST, CNRS and université Paris I, Paris. Grant from île-de-France, DIM ISC.}	
\email{mael.montevil@gmail.com}  



\keywords{Incompleteness, symmetries, randomness, critical transitions, biological evolution and ontogenesis}
\titlecomment{{\lsuper*}An introductory survey of some themes in \emph{Longo G., Montévil M.. Perspectives on Organisms: Biological Time, Symmetries and Singularities. Dordrecht: Springer, 2014.}}


\begin{abstract}
  \noindent  This short text summarizes the work in biology proposed in our book, \textit{Perspectives on Organisms}, where we analyse the unity proper to organisms by looking at it from different viewpoints. We discuss the theoretical roles of biological time, complexity, theoretical symmetries, singularities and critical transitions. We explicitly borrow from the conclusions in some key chapters and introduce them by a reflection on ``incompleteness'', also proposed in the book. We consider that incompleteness is a fundamental notion to understand the way in which we construct knowledge. Then we introduce an approach to biological dynamics where randomness is central to the theoretical determination: randomness does not oppose biological stability but contributes to it by variability, adaptation, and diversity. Thus, evolutionary and ontogenetic trajectories are continual changes of coherence structures involving symmetry changes within an ever-changing global stability.
\end{abstract}

\maketitle

\section*{Introduction} 
An analysis of biological phenomena requires many tools, thus an approach at the interface of the disciplines may help to gain insights. The construction of scientific objectivity at the core of physical theorizing is the main reference for our approach. While we borrow from the methods of physics, we do not transfer techniques and tools from that discipline, or we do not do it passively:  equations or evolution functions, for example,  are used for clarifying theoretical concepts more than for deducing and computing consequences. Thus, this short survey will not focus on the tools, but on some key conceptual constructions that frame the theoretical work in \cite{longomont}. Our hope is that this will encourage the reader to refer to our book for a more detailed discussion.

\section{A Definition of Life?}

In the multisecular debate between physicalism and vitalism, the focus has often been on the \textit{definition}  of life. A small but remarkable book by Schrödinger \cite{schrodinger} contributed to the debate in a way that we find relevant, at least in its second part which focuses on the notion of biological order. Do we provide a ``definition of life'' in our book? Do we, at least, work towards such a definition? Let's better specify how we see this issue:

\begin{description}
    \item[Primo] An ``ideal'' definition of life phenomena seems out of the question: there is no \emph{Platonic idea} of life to be grasped in a definite manner or with the maximal conceptual stability and invariance specific to mathematical notions (as there is with the definition or \emph{idea }of triangle, of Hilbert space or Turing machines \dots). It is rather a question of defining a few \emph{operational notions} enabling to draw out concepts for a systemic understanding of biological phenomena. Analogously, physics does not define ``matter'' otherwise than using operative dualities or contrapositions with the notions energy, vacuum or anti-matter, or in the opposition between fermions and bosons, for example.  Another, very rigorous, ``provable impossibility to define the object of study'' is presented in the next section. Note that Darwin's approach to evolution neither use nor need a definition of life, but needs to refer to organisms that reproduce with variations.
    \item[Segundo] The specific phenomenalities of life should be the starting point of any proposal of a operational framework. For example, it is possible that for any chosen finite list of ``defining'' properties of life, there would exist a sufficiently talented computer scientist able to create the virtual image of this property and render it on a computer screen. It is quite simple to program a virtual ``autopoietic'' system \cite{Varela1974187,varela1989} or a formalized metabolic cycle in the manner of Rosen \cite{rosen2005}  ---  see \cite{Mossio2009}, for example. However,  any human being and even non-human animals  would recognize it as a series of non-living ``virtual images'' (which are typically detectable through identical iteration, as indirectly suggested by Turing's imitation game, see \cite{turing1950,longo2008laplace}).
\end{description}

\noindent We think that the theoretical effort should focus on developing a sound intelligibility of phenomena in their constitutive, natural history. We should keep in mind the fact that \emph{any constitution is contingent}  ---  both the constitution (evolution) of life and of our historical understanding of it. That is, we stress the contingency of life phenomena and of our modest attempts to grasp its unfolding over a material evolution --- better still: over one of the \emph{possible} evolutions, taking place on \emph{this} Earth, in \emph{these} ecosystems and with \emph{this} physical matter and natural history. Our point of view includes what biologists often express when they say that nothing in biology makes sense except in the light of evolution (Darwinian and in this world) and what historians claim to be the concrete historicity of science, as a non-arbitrary, but historical tool for constructing objectivity and the very objects of knowledge.

It should be clear that we do not discuss here how ``life may have emerged from the inert'', but rather we explore how to go from the current \emph{theories }of the inert to a sufficiently robust \emph{theory} of the living. In particular, we proposed in \cite{longomont} an analysis of the specificity of the living object which may be seen as a physical singularity. We proceeded by looking first at the properties we think we need (or \emph{not}) in any theory of the ``living state of matter''. It is of course an \emph{incomplete} (see next section) attempt at providing a conceptual framework guiding more particular analyzes. 

In the following methodological reflection, we will build on the role of incompleteness in Mathematical Logic to discuss ``our theoretical endeavors towards knowledge'' (to put it in H. Weyl's words) and of its relation to conceptual or formal ``definitions'', of life in particular.

\subsection{Interfaces of Incompleteness.}

Do we need to have a definition of life to construct robust theories of the living state of matter? 
Let us now answer this question by analogy with a field where it may be dealt with in the highest rigor: Mathematical Logic. 

Is the concept of integer (thus ``standard'' or finite) number captured (defined, characterized) by the (formal) theory of numbers? Frege (1884) believed so, as the absolute concept of number was, in his view, fully characterized by Peano-Dedekind theory. In modern logical terms, we can say that Peano Arithmetic ($PA$) was ``categorical'' for Frege.  $PA$ was believed to have just one and only one model up to isomorphisms: the standard model of integers (the one which the reader learned about in elementary school, with $0$, though, and formal induction). Thus, the theory was also meant to define uniquely ``what a number is''. 

This viewpoint turned out to be blatantly wrong. Löwenheim and Skolem (1915-20) proved, by a simple proof, that $PA$ has infinitely many non-isomorphic models and, thus, that it is not categorical. Moreover, a simple theorem (``compactness'') showed that no predicate, definable in $PA$, may isolate (define) all and exactly all the standard integers (see \cite{marker2002model}). In short, any predicate valid on infinitely many standard integers must also hold for (infinitely many) non-standard integers (which cannot be considered properly ``finite'')  ---  this is also known as the ``overspill lemma''. Gödel's incompleteness theorem reinforced these negative properties: $PA$ is \emph{incomplete} or, equivalently, it has lots of logically non-equivalent models, a much stronger property than \emph{non-categoricity}. 

A fortiori, there is no hope to characterize in a finitistic way the concept of a standard (finite) integer number, or, equivalently, (Formalized) Number Theory cannot define what a number is. One has to add an axiom of infinity (Set Theory) or proper second order quantification to do so, $PA_2$, and these are infinitary or impredicative formal frames. Set Theory with an axiom of infinity and $PA_2$ are not only strict extensions but they are \emph{non-conservative} extensions of $PA$: they prove propositions of $PA$, which are unprovable in $PA$ (yet another consequence of Gödel's incompleteness). 

As a side remark, whether our theoretical proposals for biology are strict extensions of the related physical theories is surely an interesting question. However, it would be much more interesting if one of our theories or their conjunction were shown to be non-conservative with respect to a (pertinent) theory of the inert. For example, Pasteur's famous example of statistically non-balanced chirality of some macromolecules in cells is a property that can be stated in the language of physics, yet, as far as we know, it has not been derived from any physical theory. It would be fantastic if it could be justified within one of our frames, e.g.\ from a property of the phenotype at the cellular level, for example, extended criticality (see below).

In conclusion, despite its incompleteness, everybody soundly considers $PA$ as the ``natural'' (formal) theory of numbers: it elegantly singles out the main relevant, and very robust, properties of numbers ($0$, successor, induction), even though it \emph{cannot define what a number is}. There is a similarity between physics and logic.  Physics cannot define its object of study, physical matter, and Logic is another example of a sound theoretical frame, which cannot define, within itself, its object of study: the object "natural number". Moreover, we do not see a straightforward way to get out of the language of physics or of biology in the same manner that Mathematical Logic gets out of $PA$ by using infinities: what could correspond to an axiom of infinity or higher order quantification? 

We encourage the reader to pursue her theoretical work in biology without the anguishing search for a \emph{definition} of life and with the clear perspective of the intrinsic incompleteness of all our theoretical endeavors, \cite{longo2011e}. We can just hope to organize by theories some fragments of reality, whatever this latter word may mean. Let's try to do it to the best of our knowledge, in a sufficiently broad and robust way, and in full theoretical and empirical freedom. We should not necessarily feel stuck either to existing theories nor always search for the ``Ultimate (complete?) Theory'' nor the ``ultimate reduction''. As we hint in the book and several papers, molecular analyzes are not useless, of course, nor wrong a priori. In our opinion, they are just incomplete when we aim to describe phenotypes and their evolutionary and ontogenetic dynamics. 

Similarly, the issue of the emergence of life from molecules is a very relevant one.  Nevertheless, as long as we do not have a sufficiently robust, yet incomplete, theory of organisms, it is not clear what ``objects'', with what properties, should ever be shown to emerge from inert matter.

\section{Symmetry breakings and randomness} \label{sec:randosy}
Since ancient Greece (Archimedes' principle on equilibria) up to
Relativity Theory (Noether's and Weyl's work on conservation properties) and Quantum Mechanics
(from Weyl's groups to the time-charge-parity symmetry), symmetries
have provided a unified view of the principles of theoretical
intelligibility in physics. In our approach, biology requires a careful attention both to symmetries and to symmetry changes. In short, symmetry changes are symmetry breaking or formation. Symmetry changes are related with randomness and the appearance of new coherence structures, such as organisms, species, ecosystems.

In section 5 of chapter \cite{lomonsim}, we propose a preliminary and informal remark when stressing the role of randomness in biology and this remark may be significant outside of biology. Namely, we argue that every random event is associated with a symmetry change in all existing physical theories (see also \cite{longo2014}). This remark is a joint interpretation of the different frameworks for randomness that we base on the analysis of these different frameworks. In a sense its status may be compared with Church thesis.

A random event is an event where the knowledge about a system at a given time is not sufficient to deduce its future description; thus,  the event is unpredictable relatively to the intended theory.  Physics has several theoretical descriptions of random events, but in all these cases,  the description before the event determines the complete list of possible outcomes. Thus, what is unpredictable is a \emph{numerical value} in a pre-given space of observables --- modulo some finer considerations as the ones we discuss in chapter \cite{lomonphase} as for quantum field theory and statistical physics. Moreover, in physical theories, the theory provides a metric or, more generally, a measure (of probabilities or other measures) which determines the observed statistics. Then one may say that these events are random or unpredictable but only to a point:  we know the possibilities and their probability distribution. Kolmogorov's axiomatic system for probabilities works this way and provides probabilities for the possible outcomes. 

Our claim is that the various physical cases of randomness can be understood and compared in terms of symmetry breaking. 

\begin{description}
\item [Quantum Mechanics]  measurement breaks the unitarity of the quantum evolution, which amounts to say that the quantum state space assumes privileged directions (a symmetry breaking). 
 \item [Classical probabilities] the intended phase space contains the set of all possibilities. Elements of this set are symmetric in the sense that they are all possibilities. Moreover, sets of possibilities having the same probability have the same propensity to occur. Reciprocally, starting from sets that are theoretically equivalent and thus should have the same probability to occur is a common way to assign probabilities in a meaningful way. For example, the sides of a dice  are commonly assumed to be symmetric thus equiprobable and the same applies to the regions of the phase space with the same energy in the microcanonical ensemble of statistical mechanics. All these symmetries break at the occurrence of the random event, which singles out an outcome and excludes others. 
 \item [Algorithmic concurrency theory] the theory gives the possibilities (a finite list) but does not provide probabilities for them. Probabilities may be added if the physical event forcing a choice is known (but computer scientists, in programming theory and practice, usually ``do not care" --- this is the terminology they use, see \cite{longorand}). The point here is to have a program that works as intended in all cases.
 \end{description}
  
\noindent We thus related random events to symmetry breakings in the main physicomathematical frames (plus one of linguistic nature: networks' programming). In each case, we have several possible outcomes that have a symmetrical role, possibly measured by different probabilities. 
After the random event, however, one of the ``formerly possible'' situations is singled out as the actual result. Therefore, each random event that fits this description is based on a symmetry breaking, which can take different yet precise mathematical forms, depending in particular on the probability theory involved (or lack thereof). In this line of reasoning, randomness leads to a distinction between the possible and the actual result (``possible'' and ``result'' have different specific meaning depending on the theory). The symmetry is then between the different possibilities, and this symmetry breaks when we obtain one result out of them. This scheme of randomness seems quite general to us.

In the case where probabilities are defined, let us better specify the symmetries we are discussing. Let us consider an event $X$, which can be either $A$, with probability $p$  or $B$ with probability $1-p$. Then, we can consider $f_A(X)=1/p$ if $X=A$ else $f_A(X)=0$ and $f_B(X)=1/(1-p)$  if $X=B$ else $f_B(X)=0$. we see then that $f_A$ and $f_B$ have the same expectancy. It is precisely this symmetry that experimenters try to show empirically, and that legitimates the probability values.

Note that random events define a \textit{before} and an \textit{after} that the event of symmetry breaking separates. This \textit{before} and \textit{after} may be intrinsic and correspond to a genuine change of the object, for example in the case of  quantum mechanics. By contrast, it may only correspond to the knowledge of the observer, for example in the case of chaotic dynamics and we call this latter randomness epistemic.

Let us now review more closely, in a schematic way, how random events are associated to symmetry breakings:
\begin{description}
\item [Quantum Mechanics] the projection of the state vector (measurement); non-commutativity of measurement; tunneling effects; creation of a particle-antiparticle pair~\dots.
\item [Classical dynamics] the randomness associated with chaotic dynamics stems from the equivalence between different initial conditions (because the classical measurement is not pointwise), and the exponential drift of the trajectories coming from these initial conditions. 
\item[Critical transitions] the point-wise symmetry change leads to a ``choice'' of specific directions (the orientation of a magnet, the spatial orientation of a crystal, etc.). The specific directions taken are the result of fluctuations. Also, the multi-scale configuration at the critical point is random and fluctuating.
\item [Thermodynamics] the arrow of time (entropy production). This case is peculiar as randomness and symmetry breaking are not associated with an event but with the microscopic description. The time reversal symmetry breaks at the thermodynamic limit.  
\item [Algorithmic concurrency] the choice of one of the possible computational paths (backtracking is impossible).
\end{description}

\noindent If this list is exhaustive, as it seems, it is fair to say that random events, in physics, are associated with symmetry breakings (and programming follows this pattern). Note that in all these cases, one does not fit completely in our qualitative discussion and has a more complex structure: the case of thermodynamics.  Indeed, from a purely macroscopic viewpoint, there is no particular form of randomness associated with the theory, and provided that a trajectory is defined, it will be deterministic (except for critical transitions or similar situations which are discussed above).  Randomness appears at the microscopic level, either as chaotic classical dynamics or classical probabilities (in statistical mechanics). Both correspond to the analysis of their respective categories above. However, this does not explain the arrow of time, which is a particularly interesting symmetry breaking in this situation. In thermodynamics, a closed system evolves towards a maximum of entropy up to energetic constraints. This evolution is a trend towards  a symmetrization in the sense that the system evolves towards the macroscopic state to which correspond the greatest number of microscopic states (they are symmetric in the sense that they leave the macroscopic state invariant).   In this case, randomness explains the dispersion in the microscopic phase space (leading to the trend towards the macrostates corresponding to more microstates). Therefore, it is a process of symmetrization which breaks the time symmetry but does not lead to macroscopic randomness. On the opposite, it determines the macroscopic, mostly deterministic behavior of thermodynamic systems. Macroscopic randomness may still appear if there are different minima for the relevant thermodynamic potential, as in phase transitions. 

All these symmetry changes and the associated random events happen within the intended phase space, or, in other words, within the set of possibilities given by the intended physical theory and models. The challenge we are facing in biology (see \cite{lomonphase} and below), is that randomness manifests itself at the very level of the observables. Critical transitions are the closest physical phenomenon to the needs of the theoretical investigation in biology, and we will discuss them in next section.

\section{Symmetries and theoretical extensions of physical theories}

On the grounds of the previous remarks, we claim now that there are significant challenges for the proposal of mathematical and theoretical ideas in biology.
These challenges stem from  the very different roles that symmetries can play in biology when compared to physics. 

The way we picture an unifying theoretical framework
for biology is not based on specific invariants and invariants preserving transformations
  (symmetries) like in (mathematical/theoretical) physics. Instead, such a framework should
focus on the permanent qualitative changes  that modify the analysis of processes
 both in ontogenesis and evolution. Of course, these biological changes  preserve an ever changing structural stability, the coherence of organisms. 

The adaptivity of an organism and the diversity of a population are consequences of variability, thus of randomness. They contribute in an essential way to the stability of life phenomena. Thus, in a sense, variability may be considered as a primary invariant of the
living state of matter (but it is not necessarily the only one!). 

\subsection{Extended criticality}

To analyze variability, we proposed to consider
the role played by local and global symmetry changes along ``extended
critical transitions''. The notion of  phase transition was first proposed in physics by Curie, at the beginning of the last century. Phase transition typically correspond to changes of state of matter such as the transition from liquid to a gas or a solid. 
After Curie, the notion of phase transition has been deeply revised and mathematized by the introduction of renormalization methods   \cite{toulouse1977introduction,ZinnJustin_2007}. These methods are required because  certain cases, such as the paramagnetic-ferromagnetic transition,  involve a specific coherence structure  at the point of transition. This point is ``between'' two different states, a situation that typically requires the appearance of scale free patterns.  Such behavior are typical of criticality.
Critical transitions describe phase changes where a re-organization of the pertinent observables correspond to a symmetry change. In particular, a new coherence structure ``emerges'' by establishing long range correlations. Typically, the formation of a crystal or even of a snowflake, percolation, para-ferromagnetic transitions\dots may all be analyzed as critical transitions. 

We propose to analyze organisms with the tentative notion of \emph{extended} criticality, where the notion of criticality is extended from being pointwise in physics to being relevant for a whole region of the description space. Organisms are then dynamically changing
coherent structures,  global entities displaying qualitative
variability.  The coherent structure proper to critical phenomena also justifies the use of variables
depending on non-local effects. Such, an explicitly systemic approach
may help to avoid the accumulation of models and hidden
variables. In short, the notion of extended criticality provides a
conceptual framework, to be mathematized, where the dynamics of symmetries and symmetry changes provides a new, crucial role for
symmetries in biology by contrast with physics.

The concept of extended critical transition involves ubiquitous symmetry changes,  and these changes have far reaching consequences. They lead to radical methodological difficulties. In short, in  mathematics and in physics, objects are generic, they are invariants of the theory and experiment. In mathematics, a triangle, a Hilbert space\dots are used in proofs as generic, by their very definition.  Similarly, a falling object is generic: it is described as  a material point of mass $m$ at all times of its trajectory. The same applies to electrons: all electrons are assumed to obey the same ``laws''. These objects are all invariants of the theoretical and experimental frames (they are fully interchangeable, in their class). On the opposite, physical trajectories are specific, that is they are optimal in the suitable phase spaces (in the case of the electron, the ``trajectory'' is given by Schrödinger's equation thus by the trajectory of a probability amplitude in a Hilbert space). In contrast to this, we analyze biological (phylogenetic, but also ontogenetic) trajectories as generic: they are possible ones, within a phase space co-constructed by the changes of the object. Biological objects are specific in the sense that they are defined by a history and not by generic features. Two mammals have qualitative differences, and two mice also have qualitative differences. Their very names correspond to a genealogical relation, not to an identity of behaviors like in the case of the electron. Biological objects are mostly not interchangeable (or not mathematically invariant) both for the theory and for experiments (a major challenge for the interpretation of experimental results and doubly so in the case of in vitro experiments). We exemplify the instability of theoretical symmetries by a review of scale symmetries in biology \cite{scaling2014}. For example, allometry is the analysis of a quantity such as the metabolism (oxygen consumption) as a function of the size (mass) of an organism. This analysis is based on a scale symmetry, and, in mammals, the metabolism is often assumed to be proportional to $M^{3/4}$ and rhythms to $M^{1/4}$. Our analysis of the literature shows that the situation is far more complex and that some phyla have different trends: the relations above cannot be considered as a stable symmetry.

This difference between physical and biological objects  is probably the most radical change of perspective we propose. It alters the very theoretical nature of the scientific object as for proper biological observables: organisms and phenotypes. As a result, physical notions like the space of theoretical determination (phase space) cannot have the same meaning and use in biology. One of the main and maybe the main notion at the core of these changes is historicity. In evolution and development, biological objects organize themselves, and they do so in an ever changing manner, as long as their organization allow them to survive. The specificity of biological objects is associated with this historical determination and the underlying unstable mathematical symmetries. This calls for a change of perspective in the understanding of biological phenomena. Physical objects, even the most complex ones, are understood by their regularities (invariants and associated symmetries). Some physical systems are called ``historical'', for example when there are hysteresis or a few successive symmetry breaking but this historicity is limited to the state of the object and does not impact the space and the determination (equation for example) of the object like in biology.  By contrast, the most stable features of biological objects is their variability. This variability engenders diversity and contributes by this to biological structural stability, at all levels of organization. It is the reason why we put variability, understood as symmetry changes, at the core of our approach to biological phenomena.

\subsection{More on critical phase transitions in physics}
We have seen that symmetry and symmetry breaking have fundamental consequences for the determination of the behavior of objects. Theoretical symmetries correspond to conserved quantities, which are the properties of physical objects and allow their theoretical determination.

At a spontaneous symmetry breaking point, there is a loss of the structure of \emph{both} phases behaviors (the phase at the different sides of the transition). It is then logical that this loss  leads to a particular determination. More precisely, the critical point constitutes a singularity in the determination of the system because it is between two different behaviors, characterized by different relevant macroscopic phase spaces. A symmetry breaking involves the appearance of a new relevant variable describing the way in which the symmetry is broken, for example the magnetization, the structure of a crystal, etc. This variable goes from a constant zero to a finite value at the macroscopic level  which explains why the  function describing these systems cannot be analytic.

The strength of these singularities can be of different magnitudes; depending on the  Ginzburg criterion \cite{gizcri,lomoncriphy} an original method, renormalization, can be required. This criterion qualitatively assess whether averages or on the opposite fluctuations dominate a model.  In higher spatial dimensions, averages dominate since the higher the dimension of space, the more neighbors a point has. When this averaging is insufficient, renormalization methods \cite{fisher1998renormalization} are necessary to take into account the global structure of determination of the system that results from the coupling between fluctuations at all scales.

\subsection{Variability and stability}

It should be clear that when we focus on symmetry changes and variability as core notions for understanding the adaptivity and diversity proper to biological phenomena, we do not forget biological structural stability and autonomy, under ecosystemic and internal constraints. No extended criticality would ever be possible without the integrating and regulating activities proper to an organism and its relations to the ecosystem. The coherent structures characteristic of critical transitions in physics has been our initial motivation to look into criticality in the biological context.  Even though these structures change along all control parameters in a biological organism, these structures are the mathematical representation of the organismal (changing) stability: its internal and external coherence.

We have recently proposed that biological variation should be the framed by a principle \cite{chaptervariation} and that the reciprocal dependence between the parts of an organism should be described by a specific principle that we call the principle of organization \cite{chapterorganization}.

\section{Remarks on reductionism and renormalization}

In  our perspective, the peculiar phenomenality of life requires new concepts and observables. We tried to contribute to this task by the notions of extended critical transition, biological complexity, organization, proper biological time, \dots. The point is the pertinence, if any, of these treatments, ``\emph{per se}''. Those who claim that all these concepts should be reduced to  (existing?) physical theories are welcome to try: we would be very pleased and proud if the competent reductionists were able to rewrite them fully and faithfully (derive or embed them) in (existing) physical frames. However, they should first look at the history of Physics itself, where novel theoretical frames stem from the invention of new perspectives as well as new concepts and observables (inertia, gravitation, entropy, anti-matter\dots). Their pertinence had to be judged ``as such'', within their domain of meaning, not on the grounds of their reducibility to existing, thus ``safe'', explanatory grounds. 
In any cases, should reduction or unification be performed, the first question is: \emph{what theory} does one want to reduce to \emph{which theory}? Reduction, as we learn from physics and logic, is an intertheoretical issue. The case of renormalization methods exemplifies the theoretical creativity in physics and provides a very different picture than standard reductionism.

The renormalization methods are required to study critical transitions and quantum field theories where all scales contribute to a phenomenon which leads to the appearance of infinite quantities and the collapse of usual model solving.
To avoid this infinite quantities, renormalization methods use the recursive calculation of interactions on limited ranges of scales. The idea is to avoid the full set of interactions taking place at all scales and instead exhibit scaling properties asymptotically. These methods are based on the simplification of the equational with the scale change and on the stability of a part of these equations by scale change. Renormalization is useful when (a part of) the equations describing the system at different scales keeps the same form. 

The classical reductionist paradigm is to  decompose a system,  analyze the parts and (re-)compose theoretically this parts to study the system. The last part is the analysis of the interactions taking place in the system.   Renormalization methods, are outside the classical reductionist paradigm in the sense that the composition of the part is not directly solvable. Nevertheless, the intelligibility of the phenomenon still has an ``upward'' flavor in the sense that  the understanding of larger scales come from smaller scales.  The global situation may seem to be given by its (elementary) components, but the system is never understood as a combination of its parts. Renormalization analyzes a recursive sequence  of models. The ``locus of the objectivity'' is not in the description of the parts but in the stability of the equational determination when taking more and more interactions into account. This rationale also holds for those critical phenomena where some parts, atoms, for example, can be objectivated extrinsically to the renormalization and have a characteristic scale. In general, only scale invariance matters and the contingent choice of a fundamental (atomic) scale is irrelevant. Actually, in quantum fields theories, there is no known relevant elementary scale.  Again, such a scale would not play a significant role since the objectivity of the approach lies in its inter-scale relationships, see for example \cite{ZinnJustin_2007} for a technical discussion.

In short, even in physics, there are situations where the whole is not the sum of the parts because the parts cannot be summed on. This issue is not unique to quantum fields as it is also relevant for classical fields. In all these situations, the intelligibility is obtained by a scale symmetry. This symmetry is why choices of  fundamental scales  are arbitrary for the models of these phenomena, see \cite{Longo_2012_From} for further discussions.

Broadly speaking, the theoretical principles that we propose in \cite{longomont} constitute an \emph{extension} of existing physical theories since they address observables and quantities unique to life phenomena. They preserve the same formal mathematical structure and, if we set the value of the considered observables or parameters to $0$, they lead us back to the case of the inert. That is, if there is no protention \cite{lomonproret}, no second temporal dimension \cite{lomongeo}, no extension of criticality \cite{lomonextend}, zero anti-entropy \cite{lomonanti}, one returns to physical frames. Our theoretical frameworks are thus compatible, although they may be irreducible to ``existing physical theories''. That is, they are reducible to physics \emph{as soon as} they are outside the extended critical zone having its own temporality and its anti-entropy, or as soon as these specific quantities go to $0$.

In the next section, we will explore the consequences of our analysis on the notion of phase spaces, discuss causality and introduce the concept of enablement.

\section{Phase spaces and enablement}
 
We have  discussed the role of invariance, symmetries and conservation properties in physical theories, as presented in chapter \cite{lomonsim}. 
Our aim, here, following chapter \cite{lomonphase}, is to hint that the powerful methods of physics do not apply as such to biology.  More precisely, physical methodology pre-defines phase spaces on the grounds of the observables and the invariants in the ``trajectories'' (the symmetries of the equations) and we argue that this methodology has to be reevaluated in biology.

In biology, symmetries at the phenotypic level are continually changed, beginning with cell proliferation, up to the ``structural bifurcations'' which yield speciations in evolution. Thus, there are no biological symmetries that are \emph{a priori} preserved except   for some time and we call these symmetries and invariants constraints. There are no sufficiently stable mathematical regularities and transformations to allow an equational and law like description entailing the phylogenetic and ontogenetic trajectories. Biological changes involve cascades of symmetry changes and thus  cumulative historical dynamics. Each symmetry change is associated with a random event (quantum, classical or due to bio-resonance, see \cite{buiatti2011randomness}), while the global shaping of the trajectory, by selection say, is also due to non-random events. In this sense biological trajectories are generic: they are just 
possible ones and yield a historical result, that is an individuated, specific organism (see  \cite{bailly2011,longo2011c,lomonextend}).

In other words, this sum of individuals and individualizing histories, co-constituted within an ever-changing ecosystem, does not allow a compressed, finite or formal description of the space of possibilities. The  actual biological phase space (functions, phenotypes, organisms) is not described by a definite axiomatic. Biological  possibilities are  the result of an unpredictable sequence of symmetry changes. This situation is in contrast to the invariant (conservation) properties which determine physical ``trajectories'', in the broad sense (including for Hilbert's spaces, in Quantum Mechanics). 

An immense literature has been tackling ``emergence'' in life phenomena. In the technical analyzes, the strong and dominating theoretical frames inherited from mathematical physics (or even computing) remain the main reference. From Artificial Life to Cellular Automata and various very rich analysis of dynamical systems, the space for intelligibility is given  \emph{a priori}. It takes  the form  of one or more pre-defined phase spaces, possibly to be combined by adequate mathematical forms of products (Cartesian, tensorial products \dots). A very rich and motivated framework for these perspectives is summarized in \cite{rooney2007energetics}. Well beyond the many analysis which deal with equilibrium systems, an inadequate frame for biology, these authors analyze interactions between multiple attractors in dissipative dynamical systems, possibly given in two or more phase spaces (the notion of attractor is a beautiful mathematical notion, which requires explicit equations 
or evolution functions --- solutions with no equations --- in pertinent phase spaces). Then, two or more deterministic, yet highly unpredictable and independent systems, interacting in the attractor space, may ``produce persistent attractors that are offsprings of the parents\dots. Emergence, in this case, has a precise meaning because no trajectories exist linking the child to either parent (p. 158) \dots [The] source [of emergence] is the creation, evolution, destruction, and interaction of dynamical attractors (p. 179)''.

This analysis is compatible with ours, and it may enrich it by a further component, in pre-given interacting phase spaces. Yet, we go somewhat beyond pre-given phase spaces, from a critical perspective, which, per se, is a tool for intelligibility. Below, we will hint again to further possible (and positive) work, besides negating the possibility of an \textit{a priori} and compressed mathematical description of (combined) spaces of evolution.

In summary, in our approach, the intrinsic unpredictability of the very \emph{Phase Space} of phylogenetic (and ontogenetic) dynamics corresponds to:

\begin{enumerate}
\item physical and properly biological randomness. In particular,  bio-resonance  is due to interacting levels of organization, as a component both of integration and regulation in an organism. This includes the  amplification of random fluctuations in one level of organization through the others;
\item extended criticality, as a locus for the correlation between  symmetry breaking and randomness;
\item cascades of symmetry changes in (onto-)phylogenetic trajectories;
\item enablement, or the co-constitution of niches and phenotypes, a notion to be added to the physical determination and that we define below.
\end{enumerate}

\noindent By the lack of mathematically stable invariants (stable symmetries), there are no laws that entail, as in physics, the biological observables in the becoming of the biosphere. In physics, the geodetic principle mathematically forces objects never to go wrong. A falling stone follows exactly the gravitational arrow. A river goes along the shortest path to the sea, and it may change its path by nonlinear well definable interactions as mentioned above, but it will never go wrong. These are all optimal trajectories. Even though it may be very hard or impossible to compute them, they are unique, by principle, in physics. Living entities, instead, may follow many possible paths, and they go wrong most of the time. Most species are extinct, almost half of fecundations in mammals do not lead to a birth, and an amoeba does not follow, exactly, a curving gradient --- by retention it would first go on the initial tangent, then it corrects the trajectory, in a protensive action. In short, life goes wrong most of the time, but 
it 
``adjusts'' to the environment and may change the environment \cite{nicheconstr}: it is adaptive. It maintains itself, always in a critical transition, that is within an extend critical interval, whose limits are the edge of death. It does so by changing the observables, the phenotypes, and its niche --- in the sense of Darwinian correlated variations of organisms and ecosystems. Thus, it is the very nature and phase space of the living object that changes, in contrast to physics. 
 
We must ask new scientific questions and invent new tools to understand these co-constitutions that is to say the way  organisms co-evolve and make their worlds together. We consider this feature as a central component of the biosphere's dynamics.  
The instability of theoretical symmetries in biology is not, of course, the end of science, but it sets the limits of the transfer of physicomathematical   philosophy and methods to biology. As such, the instability of theoretical symmetries in biology can be considered a ``negative result''. Kant already doubted of the applicability of physicomathematical reasoning to biology, \cite{kant1781critique}. In biological evolution, we cannot use the same very rich interaction with mathematics than in the core of physical theories. However, mathematics is an adaptive human construction: an intense dialogue with biology may shape  new scientific paths, concepts, structures, as it did with physics since Newton. 

By providing some theoretical arguments that yield this ``negative result'' in terms of symmetries and critical transitions, we hope to  provide also some tools for a new opening. Negative results  marked the beginning of new sciences in several occasions: the thermodynamic limit to energy transformation (increasing entropy), Poincaré's negative result (as he called his Three Body Theorem), Gödel's theorem (which set a new start to Recursion Theory and Proof Theory) all opened new ways of thinking, \cite{longo2012c}. Limits clarify the feasible and the nonfeasible with the existing tools and may show new directions by their very nature if these limits have a sufficiently precise, scientific content.

The scientific answer we propose to this end of the physicalist certitudes is based on our analysis of symmetry changes in extended critical transitions and the notion of ``enablement'' in evolution (and ontogenesis). Enablement is a form of causation that is proper to biology and has never been developed in physics. A biological trait enables the appearance of novel traits that do not result from the properties of the initial trait. Instead, any biological trait have a specific form of causal power: it makes new traits possible and these new possibilities cannot be predicted on the basis of the current state of affairs. Enablement concerns how organisms co-create their worlds, with their changing symmetries and coherence structures, such that they can exist in a qualitatively expanding universe.

Our thesis is that evolution  and ontogenesis  are  ``diachronic
processes'' of becoming that ``enable'' the future state of affairs
and  do not cause it in the physical sense.  Reproduction with
variation, beginning with the cellular level, is the primary dynamics
(the ``default state'', see \cite{solono16}).  Moreover, Galileo and Newton's mathematization of trajectories concerns only Aristotle's ``efficient cause''. Instead,  such \emph{entailed causal relations must be enriched by ``enablement'' relations} for biological processes. Physical quantities typically play a different role in biology than in physics. In biology, we consider that they play the role of constraints, limiting possibilities on the one side and enabling behaviors on the other side.

Life is caught in a causal web but also lives in a web of enablement and radical emergence of life from life, whose intelligibility may  largely be given in terms of symmetry changes and their association to random events at all levels of organization. 

Enablement is crucial to understand life persistence. Variability, thus  diversity and adaptability are an integral component of life persistence. Our theoretical frame, in particular, is based on reproduction with variation and motility as the proper default state for the analysis of phyllo- and ontogenesis \cite{chapterconstraints} where selection shapes the bubbling forth of life by excluding the incompatible. 

As hinted in section 5 of \cite{lomonphase}, a long term project would be to better quantify our approaches to the two-dimensional time for rhythms, to extended criticality and to anti-entropy (basically an evaluation of biological complexity, see chapter \cite{lomonanti}). This would allow to construct an abstract phase space based on these mathematically stable properties. The analysis should follow the nature of Darwin's evolution, which is a historical science, not meant to ``predict'' yet giving remarkable insights on the living. Thus, the dynamics of extended criticality or anti-entropy should just provide the evolution of these state functions, or how these abstract observables may develop with respect to the intended parameters and over time. And this, without being ``projectable'' on specific phenotypes, even not in probabilities, as it is instead possible for Schrödinger's state functions in Quantum Mechanics. To this purpose, one should give a biologically interesting measure for extended criticality and describe it in a 
quantitative way in the abstract space of extended critical transitions, that is to say the qualitative evolution of life. In a preliminary way, we have been able to do so, by following Gould's analysis of increasing biological complexity by analyzing the evolutionary dynamics of a global observable we  call anti-entropy, \cite{lomonanti}.

\section*{Acknowledgment}
  The authors wish to acknowledge our preliminary joint work with Francis Bailly and many fruitful discussions with Carlos Sonnenschein, Ana Soto, Matteo Mossio, Arnaud Pocheville, Stuart Kauffman.

\bibliographystyle{alpha}
\bibliography{bib}

\end{document}